\documentclass[twocolumn, aps,showpacs,amsmath,reprint,amssymb,amsart,superscriptaddress,prl]{revtex4-1}

\usepackage[FIGTOPCAP,tight]{subfigure} 
\usepackage{amsmath}
\usepackage{amssymb}
\usepackage{amsthm}
\usepackage{float}
\usepackage[T1]{fontenc}
\usepackage[latin9]{inputenc}
\usepackage{geometry}
\usepackage{graphicx}
\usepackage{esint}
\usepackage{array}
\usepackage{multirow}
\usepackage{color}
\usepackage[normalem]{ulem}

\newcommand{\ket}[1]{\ensuremath{\left|#1\right\rangle}}

\begin{document}

\title{Optical sectioning in induced coherence tomography with frequency-entangled photons}

\author{Adam Vall\'{e}s} \email{adam.valles@icfo.eu; Present address: School of Physics, University of the Witwatersrand, Private Bag 3, Johannesburg 2050, South Africa}
\affiliation{ICFO-Institut de Ciencies Fotoniques, Barcelona
Institute of Science and Technology, Mediterranean Technology
Park, 08860 Castelldefels, Barcelona, Spain}

\author{Gerard Jim\'{e}nez}
\affiliation{ICFO-Institut de Ciencies Fotoniques, Barcelona
Institute of Science and Technology, Mediterranean Technology
Park, 08860 Castelldefels, Barcelona, Spain}

\author{Luis Jos\'{e} Salazar-Serrano}
\affiliation{ICFO-Institut de Ciencies Fotoniques, Barcelona
Institute of Science and Technology, Mediterranean Technology
Park, 08860 Castelldefels, Barcelona, Spain}

\author{Juan P. Torres} \email{juanp.torres@icfo.eu}
\affiliation{ICFO-Institut de Ciencies Fotoniques, Barcelona
Institute of Science and Technology, Mediterranean Technology
Park, 08860 Castelldefels, Barcelona, Spain}
\affiliation{Department of Signal Theory and Communications,
Universitat Politecnica de Catalunya, Campus Nord D3, 08034
Barcelona, Spain}

\date{\today}

\begin{abstract}
We demonstrate a different scheme to perform
optical sectioning of a sample based on the concept of induced
coherence [Zou et al., {\em Phys. Rev. Lett.} \textbf{67}, 318
(1991)]. This can be viewed as a different type of optical coherence
tomography scheme where the varying reflectivity
of the sample along the direction of propagation of an optical
beam translates into changes of the degree of first-order
coherence between two beams. As a practical advantage the scheme
allows probing the sample with one wavelength and measuring photons
with another wavelength. In a bio-imaging scenario, this would
result in a deeper penetration into the sample because of probing
with longer wavelengths, while still using the optimum wavelength
for detection. The scheme proposed here could achieve submicron axial resolution by making use of nonlinear parametric sources with broad spectral bandwidth emission.
\end{abstract}


\maketitle

\section{Introduction}

In 1991, Huang et al.~\cite{huang1991}
put forward a three-dimensional noninvasive optical imaging technique that
permits cross-sectional and axial high-resolution tomographic
imaging of biological tissue. They named the new technique optical
coherence tomography (OCT) and demonstrated it obtaining
high-resolution images of the layers that make up the retina. The
axial and transverse resolutions of OCT are independent. To obtain
information in the transverse direction (plane perpendicular to
the beam propagation), OCT focuses light to a small spot that is scanned over
the sample. To obtain information in the axial direction (along
the beam propagation), OCT uses a source of light with short coherence
length that allows optical sectioning of the sample.

In the same year, Zou et al.~\cite{zou1991} introduced the
concept of {\em induced coherence}. When two second-order
nonlinear crystals (NLC$_1$ and NLC$_2$) are optically pumped by a
coherent wave, a pair of signal and idler photons
might emerge (signal $s_1$ and idler $i_1$ from NLC$_1$; signal
$s_2$ and idler $i_2$ from NLC$_2$) by means of the nonlinear
process of parametric down-conversion. Most experiments are
usually done in the regime of very low parametric
gain (weak pumping), so that paired photons are
expected to be generated in one or the other
crystal~\cite{Mandel1995Optical}.

In the absence of any other injected signal or idler beams, the
two signal beams show no first-order coherence
($|g^{(1)}_{s_1,s_2}|=0$)~\cite{firstorder,glauber1963} and thus
do not give rise to interference when recombined in a beam
splitter~\cite{ou1990}. However, if idler $i_1$ is injected into
the second nonlinear crystal and the experimental configuration is
designed to make idlers $i_1$ and $i_2$ indistinguishable after
NLC$_2$, the signal photons $s_1$ and $s_2$ will show first-order
coherence, i.e., $|g^{(1)}_{s_1,s_2}|=1$. If idler $i_1$ traverses
an element with reflectivity $\tau$ before impinging on NLC$_2$ (or transmissivity $\mu$ depending on the setup configuration),
there is a loss of first-order coherence between the signal
photons coming out from the two nonlinear crystals. This effect
could be observed in the temporal~\cite{zou1991} and frequency
domains~\cite{zou1992}, and it should still be present in the case
of strong pumping ~\cite{belinsky1992,wiseman2000}.

There has been a growing interest in recent years in these
so-called {\em nonlinear interferometers} \cite{chekhova2016}, not
only because of their importance to unveil the interplay between
information and coherence in quantum theory
\cite{heuer2015a,heuer2015b}, but also because of their
applications in quantum information and quantum metrology. For
instance, Kalashnikov et al. \cite{Kalashnikov2016} showed that a
nonlinear interferometer allows performing spectral measurements
in the infrared range using visible-spectral-range components,
avoiding the use of optical equipment suited for operation in the
infrared range that may have inferior performance and higher
cost.

Recently, Barreto et al.~\cite{barreto2014,hochrainer2017} used
the concept of induced coherence to demonstrate a
two-dimensional quantum imaging system, where photons used to illuminate the
object do not have to be detected at all, which enables the probe
wavelength to be chosen in a range for which suitable detectors
are not available. We might call the imaging system {\em
induced coherence tomography}.

Here we go one step further and demonstrate in a {\em
proof-of-concept} experiment that a nonlinear interferometer based
on the concept of induced coherence can be used to perform three-dimensional
imaging of a sample, i.e., in addition to obtaining information in
the transverse plane (plane perpendicular to the beam propagation), which was
demonstrated in Ref.~\cite{barreto2014}, it can also provide optical
sectioning of the sample (information in the axial direction,
along the beam propagation), which we demonstrate here.

\begin{figure}[t]
\centering
\includegraphics[width=8.6cm]{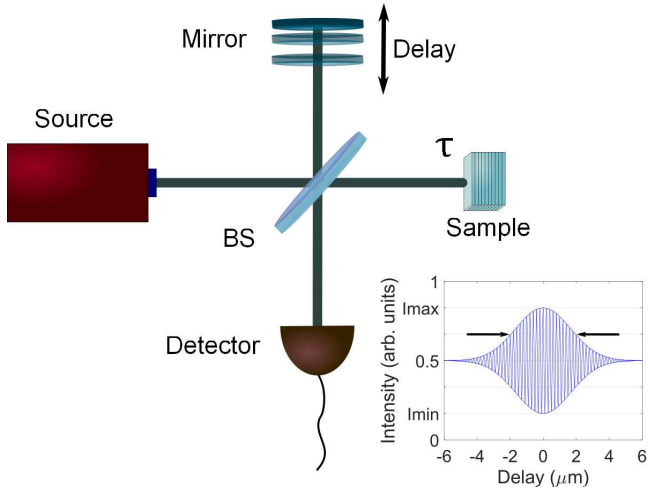}
\caption{Sketch of a time-domain OCT scheme and typical
interferogram obtained for a certain axial measurement. The
interferogram shows the effect on axial resolution of the
low-coherence length of the source of light.} \label{StandardOCT}
\end{figure}

\begin{figure}[t]
\centering
\includegraphics[width=8.6cm]{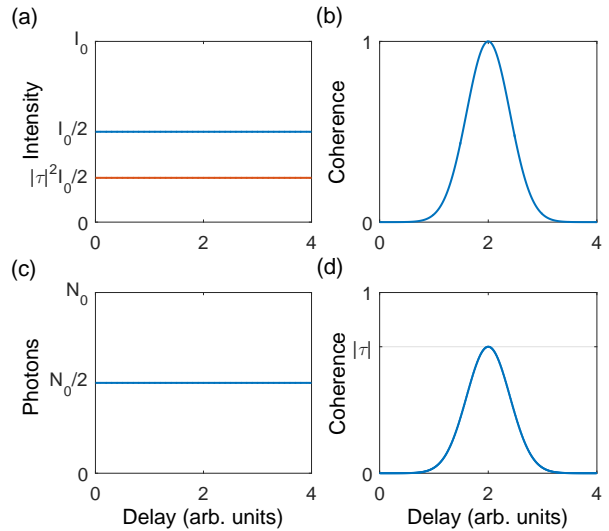}
\caption{Differences between a simplified but {\em typical} OCT
scheme and the new configuration are demonstrated. (a, c)
Intensity (or photon flux) traversing the reference and sample
arms of the interferometer. (a) In OCT, the intensity of the
reference beam is $I_0/2$, and the intensity coming from the
sample is $|\tau|^2I_0/2$. $I_0$ is the total intensity. (c) In
our optical sectioning scheme, in the very low parametric gain
regime, the two signal beams traversing each arm of the
interferometer contain $N_0/2$ photons, independently of the
reflectivity $\tau$. $N_0$ is the total number of photons
propagating through the interferometer. (b, d): Degree of
first-order coherence between light beams propagating in the two
arms of the interferometer. (b) Coherence in OCT, and (d)
coherence in our scheme.} \label{IoNoCohTheory}
\end{figure}

In doing this, we put forward indeed a new type of OCT scheme
based, however, on a different physical principle. This is explained
with the help of Figs.~\ref{StandardOCT} and ~\ref{IoNoCohTheory}.
In OCT, different layers of the sample show different reflectivity
($\tau$) or backscattering. For each axial measurement, the wave
reflected from the sample and the reference wave carry different
intensities, $|\tau|^2I_0/2$ and $I_0/2$, respectively [see
Fig.~\ref{IoNoCohTheory}(a)]. They show first-order coherence only
for a given delay [see Fig.~\ref{IoNoCohTheory}(b)] and interfere
for this delay with a visibility that depends on $\tau$. OCT
performs thus direct measurements of the reflectivity; it {\em
does not measure first-order coherence} as the name of the
technique might wrongly lead one to think. The low-coherence length
of the source provides positioning of the reflectivity
measurement, the exact depth into the sample that is being
analyzed.

On the other hand, in the scheme based on induced coherence in the
very low parametric gain regime, the flux of photons in the two
arms of the interferometer is the same ($N_0/2$) [see
Fig.~\ref{IoNoCohTheory}(c)]. However, there is a loss of
coherence between both beams [see Fig.~\ref{IoNoCohTheory}(d)]
that is due to the reflectivity of the sample (see
the Appendix). Therefore, and contrary to
common OCT configurations, first-order coherence
plays a double role in induced coherence tomography: i) it carries
the sought-after information about the reflectivity of the sample
and ii) it provides axial optical sectioning of the sample.

An OCT scheme with the word {\em quantum} attached to it was
demonstrated some years ago \cite{nasr2003,abouraddy2002}. It
showed, as a characteristic element, certain immunity to the
presence of depth-dependent dispersion in the sample that
deteriorates the resolution achievable in an OCT scheme
\cite{wang2003}. This dispersion cancellation effect also appears
when using phase-sensitive cross-correlated beams that, however,
show no entanglement \cite{erkmen2006,gouet2010}. This so-called
quantum OCT scheme is fundamentally different from ours in two
important aspects: first, they use entangled photons that,
however, are not embedded in a nonlinear interferometer. Second,
their OCT scheme is based on the measurement of second-order
coherence functions (coincidence counts measurements), while our
scheme makes use of first-order coherence functions, as is the
case of common OCT schemes.

\section{Brief description of the concept of induced coherence}
\begin{figure}[t!]
\centering
\includegraphics[width=8.6 cm]{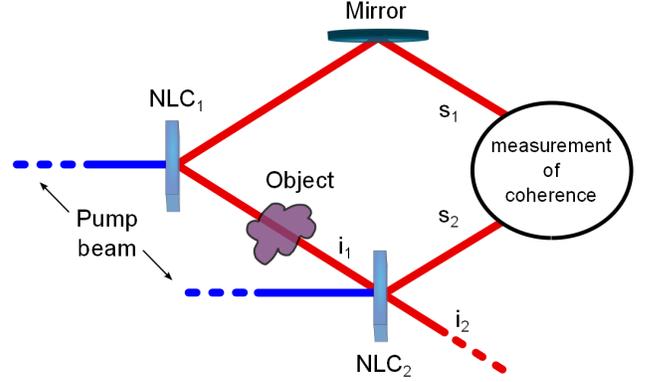}
\caption{Simplified sketch of a nonlinear interferometer aimed at
introducing the concept of induced coherence by using two nonlinear crystals (NLC$_{1\&2}$).} \label{ZWM91Setup}
\end{figure}

Since the concept of induced coherence is the physical basis that
makes possible the results presented in this paper, for the sake of
clarity we briefly describe here the concept (see Fig.~\ref{ZWM91Setup}). For
simplicity, we consider the single-mode case in this section.
A more detailed description of the multimode (multifrequency) case is
considered in the Appendix.

Two second-order nonlinear crystals are coherently pumped by a
strong pump beam. In the first nonlinear crystal (NLC$_1$), pairs of signal
(annihilation operator $a_{s_1}$) and idler ($a_{i_1}$) photons
are generated by means of the spontaneous parametric
down-conversion process (SPDC). The relationship between the input ($b_s$ and
$b_i$) and output operators ($a_{s_1}$ and $a_{i_1}$) can be
described by a Bogoliuvov transformation ~\cite{navez2001,brambilla2004,torres2011}:
\begin{eqnarray}
& & a_{s_1}=U b_{s}+V b_{i}^{\dagger}, \nonumber \\
& & a_{i_1}=U b_{i}+V b_{s}^{\dagger},
\end{eqnarray}
where $|U|^2-|V|^2=1$. The idler photon traverses a lossy object
with transmissivity $\mu$:
\begin{equation}
a_{i_1} \Longrightarrow \mu a_{i_1}+f,
\end{equation}
being $[f,f^{\dagger}]=1-|\mu|^2$ \cite{haus2000,boyd2008}.

After traversing the lossy object, $i_1$ enters the second
nonlinear crystal (NLC$_2$). The annihilation operator of signal $s_2$ after
the second nonlinear crystal is
\begin{eqnarray}
& & a_{s_2}=U c_{s}+V \left[ \mu^* a_{i}^{\dagger} + f^{\dagger} \right] \nonumber \\
& & \Longrightarrow a_{s_2}=U c_{s}+|V|^2 \mu^* b_{s} + V U^{*}
\mu^{*} b_i^{\dagger}+V f^{\dagger},
\end{eqnarray}
where $c_s$ is the input signal operator in the second nonlinear
crystal.

The degree of coherence between the two signal waves, $s_1$ and
$s_2$, can be quantified by the normalized degree of first-order
coherence, which reads
\begin{equation}
g_{s_1,s_2}^{(1)}=\frac{\langle a_{s_1}^{\dagger}
a_{s_2}\rangle}{\sqrt{\langle a_{s_1}^{\dagger} a_{s_1} \rangle}
\sqrt{\langle a_{s_2}^{\dagger} a_{s_2} \rangle}}.
\end{equation}

If the quantum states corresponding to the operators $b_s$, $b_i$
and $c_s$ are the vacuum state, the flux rate of $s_1$ photons is
$\langle a_{s_1}^{\dagger} a_{s_1} \rangle=|V|^2$ and the flux
rate of $s_2$ photons is $\langle a_{s_2}^{\dagger} a_{s_2}
\rangle=|V|^2 [1+|\mu|^2 |V|^2]$. We can obtain~\cite{belinsky1992,wiseman2000}:
\begin{equation}
|g_{s_1,s_2}^{(1)}|=|\mu| \sqrt{\frac{1+|V|^2}{1+|\mu|^2|V|^2}}.
\end{equation}

If $|\mu|=1$ there is first-order coherence between the two
signals. Injection of $i_1$ into the second nonlinear induces
coherence between signals $s_1$ and $s_2$. Notice that it is
important that idler $i_2$ is indistinguishable from idler $i_1$
after the second nonlinear crystal. This is why sometimes the
induced coherence is related to the indistinguishability of the
idler waves.

If $|\mu|=0$, there is no first-order coherence between signals,
since
\begin{eqnarray}
& & a_{s_1}=U b_{s}+ V b_{i}^{\dagger}, \nonumber \\
& & a_{s_2}=U c_{s}+ V f^{\dagger},
\end{eqnarray}
and $\langle b_s^{\dagger} c_s\rangle=0$ and $\langle b_i
f^{\dagger}\rangle=0$. In this case, we have two independent
spontaneous parametric down-conversion processes, therefore the
resulting signal waves show no coherence. In an intermediate case,
there is partial coherence between the two signal waves.

Under most circumstances, as is the case here, experiments work
in the low parametric gain regime, where $V$ is extremelly small
($|V| \ll 1 $), so $|U| \sim 1$. In this case, we have
\begin{equation}
|g_{s_1,s_2}^{(1)}|=|\mu|.
\end{equation}

In this scenario a pair of signal-idler photons is generated in
one crystal or the other, since the probability to generate two
pairs of signal-idler photons is extremely low.

\begin{figure*}[t]
\centering
\includegraphics[width=15cm]{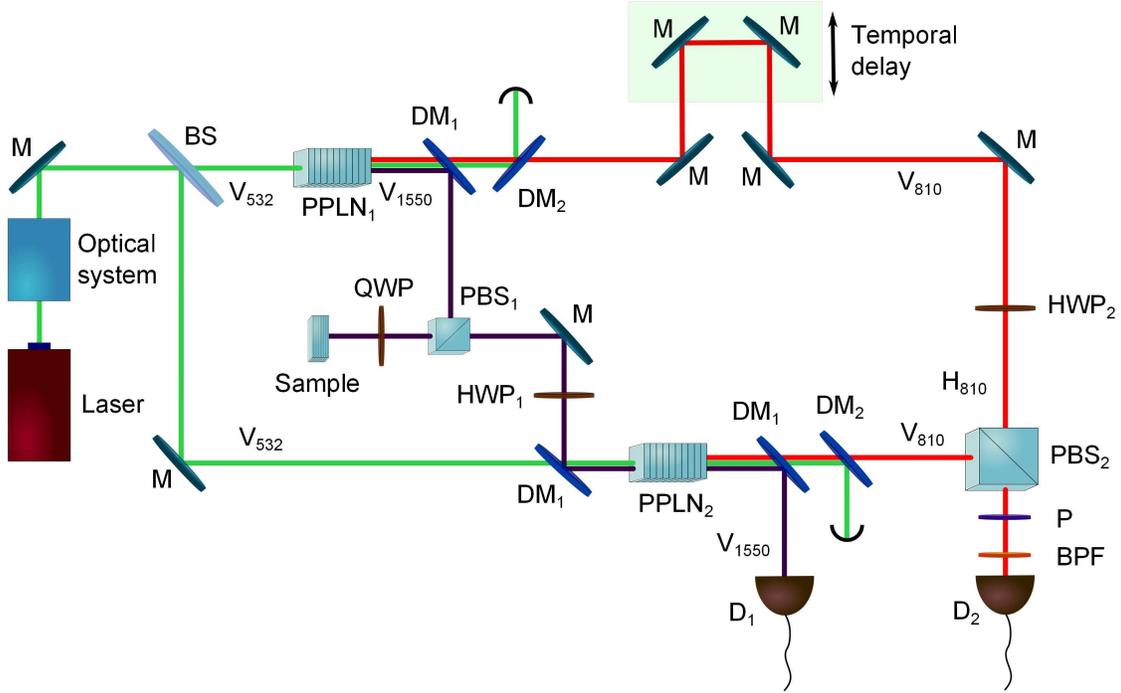}
\caption{Experimental set-up for observing optical sectioning
based on the concept of induced coherence. Laser: Verdi V10;
Optical system: linear attenuator and short-pass filter; BS: beam
splitter for the pump beam; PPLN$_{1\&2}$: periodically polled
lithium niobate (nonlinear crystals); DM$_{1\&2}$: dichroic
mirrors; Sample: mirror and a variable neutral density filter;
Temporal delay: 6-mm-long stepper motor; QWP: quarter-wave plate;
HWP$_{1\&2}$: half-wave plates; D$_1$: optical spectrum analyzer
(OSA) at the telecom wavelength; D$_2$: single-photon counting
module; M: mirrors; P: polarizer; BPF: band-pass filter; H and V: horizontal and vertical polarizations, respectively, and
the sub-index indicates the wavelength of the beam. For the sake
of simplicity, focal distances and position of lenses are not
shown. However, the exact position of the lenses in the lower
interference arm before the PPLN$_{2}$ has an important role when
the distinguishability between the two idler spatial modes is at
stake.} \label{iOCTSetup}
\end{figure*}

\section{Experimental set-up}

Figure~\ref{iOCTSetup} depicts the experimental set-up. The laser
that pumps the two nonlinear crystals is a high-power
continuous-wave (CW) Verdi V10 (Coherent, wavelength of 532 nm). The path difference traveled by the pump beam in
its way towards the two nonlinear crystals should be smaller than
the coherence length of the pump beam to allow interference
between the down-converted photons \cite{heuer2014}. The pump
beam is split with a 50:50 beam splitter (BS), so that the same
pump power impinges on two periodically polled lithium niobate
crystals (PPLN$_1$ and PPLN$_2$). These crystals mediate the
absorption of a 532 nm pump photon and the generation of two
lower-frequency photons, signal and idler, by means of SPDC. The
process is nondegenerate type-0, meaning that all three photons,
pump, signal and idler, have the same vertical polarization.
Signal and idler photons are generated with different central
wavelengths, 810 nm and 1550 nm, respectively. The efficiency of
the SPDC process is extremely low, so we can neglect the
probability to generate two pairs of signal-idler photons, each
pair in a different crystal, at the same time.

\begin{figure}[t]
\centering
\includegraphics[width=8.6cm]{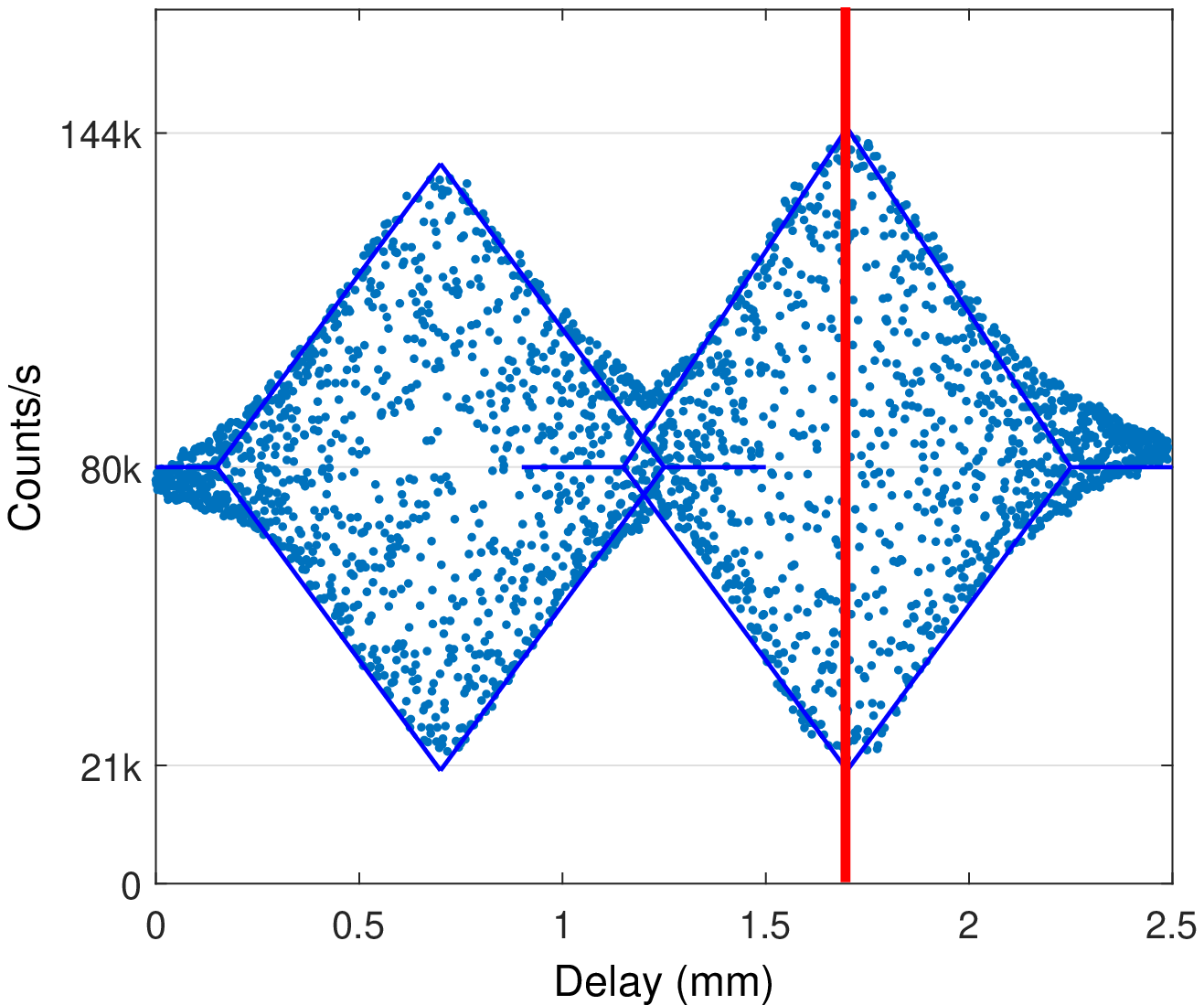}
\caption{Degree of first-order coherence measured from two layers
separated 1 mm apart. We detect 809.4 nm photons with diagonal
polarization at the output of PBS$_2$, changing the path
difference between the two arms of the interferometer in
micrometric steps ($1 \, \mu$m) for every point depicted. We
obtain a maximum interference visibility of $V=73\%$, within the
region marked in red. Blue dots correspond to experimental
data, and the solid curve stands for the theoretical prediction
given by Eq.~(\ref{g1}), taking into account our given visibility values.} \label{PolCohLength}
\end{figure}

Signal and idler photons leaving PPLN$_1$ are separated by a
dichroic mirror (DM$_1$). The 810 nm signal photon is transmitted,
forming the upper arm of the Mach-Zehnder interferometer. Its
polarization is changed with the help of the half-wave plate
(HWP$_2$) and joins the signal photon coming from PPLN$_2$ at the
polarization beam splitter (PBS$_2$). Both signal photons have
orthogonal polarizations before reaching PBS$_2$. The measurement
is carried out by detecting the polarization state of the 810 nm
signal photons after PBS$_2$. If there is coherence between signal
photons ($|\tau|=1$) the polarization state after PBS$_2$ will be
\begin{equation}
\ket{\Psi}=\frac{1}{\sqrt{2}}\{\ket{H} + \exp{(i\phi)}\ket{V}\},\label{Psi}
\end{equation}
where $\phi$ is the phase difference between the two interfering
arms, introduced by the temporal delay stage. Whereas if there is no polarization coherence ($|\tau|=0$), the
polarization state will read
\begin{equation}
\rho=\frac{1}{2}\{\ket{H}\langle H| + \ket{V} \langle
V|\};\label{rho}
\end{equation}
therefore there will not be any phase difference dependence.

The 1550 nm idler photon coming from PPLN$_1$ is
the one that interacts with the sample. It is reflected in the
dichroic mirror DM$_1$, starting the lower arm of the
interferometer. It is reflected again in the polarization beam
splitter (PBS$_1$), due to its vertical linear polarization. It
traverses a quarter-wave plate (QWP) that changes its polarization
to circular. It interacts with the sample, formed by a tunable
neutral density filter and a movable mirror that can be displaced
longitudinally up to by 1 mm. This is equivalent to the presence
of a layer of that thickness for low reflectivity values. The
mirror reflects back the idler photon with probability $|\tau|^2$.
The QWP changes its polarization to horizontal. This photon, now
carrying the information of the sample ($\tau$), is transmitted
through PBS$_1$ due to its horizontal polarization. Later it is
rotated again to vertical polarization with a half-wave plate
(HWP$_1$) in order to erase all distinguishing information with
respect to the second 1550 nm idler photon. With another dichroic
mirror DM$_1$, the 1550 nm idler photon that is generated in the
first nonlinear crystal and probed the sample, overlaps spatially
with the pump beam that illuminates the second nonlinear crystal,
and consequently also with the second 1550 nm idler photon.

After the second nonlinear crystal (PPLN$_2$), the second signal
photon is separated from the two spatially overlapping idler
photons, which are reflected in DM$_1$ and coupled into a
single-mode fiber for alignment purposes. The second 810 nm signal
photon continues the lower interferometer arm until it reaches
PBS$_2$. A temporal delay is implemented in the upper arm of the
interferometer, formed by two mirrors on top of a platform that
can move in steps of the order of tens of nanometers because of a
6-mm stepper motor (Thorlabs Z806) attached to it.

The two 20-mm-long PPLN crystals are mounted on top of ovens
(Covesion), being able to adjust their temperature by a tenth of
degrees Celsius. This change of temperature induces a variation in
the spectral response of the nonlinear crystals, leading to
different phase-matching conditions for each temperature. In order
to oversee the spectral overlap between idler photons originated
in the two PPLN crystals, its spectrum is measured with an optical
spectrum analyzer (OSA). Notwithstanding, the detection of the
1550 nm photons is not necessary for the correct functioning of
the OCT scheme, its detection serves only for monitoring and
alignment purposes.

The pump beam at 532 nm is separated from the signal after being
reflected in the dichroic mirror DM$_2$. The residual pump power
still existing at the output of PBS$_2$, overlapping with the two
orthogonally polarized 810 nm signal photons, is filtered out by a
band-pass filter (BPF). We  also implement a polarizer (P) that
project the incoming signal photons into the polarization diagonal
state
\begin{equation}
\ket{D}=\frac{1}{\sqrt{2}}\{\ket{H}+\ket{V}\},
\end{equation}
being able to measure the phase dependence given in
Eq.~(\ref{Psi}). Finally, the interference signal is coupled into
a single-mode fiber and measured with a silicon-based single
photon detector (Perkin-Elmer).

The results presented in the next section are interferometric
measurements, and for the sake of clarity, we summarize here what
constitutes the interferometer. The Mach-Zehnder interferometer
starts in PPLN$_1$. The upper arm is formed by the 810 nm signal
photon generated in the first nonlinear crystal until it reaches
the polarization beam splitter PBS$_2$. The lower arm is formed by
the 1550 nm idler photon generated also in the first nonlinear
crystal, until it reaches the second nonlinear crystal (PPLN$_2$).
It continues with the 810 nm signal photon generated in the second
nonlinear crystal until it reaches the other input port of the
polarization beam splitter PBS$_2$.

\begin{figure}[t!]
\centering
\includegraphics[width=8.6cm]{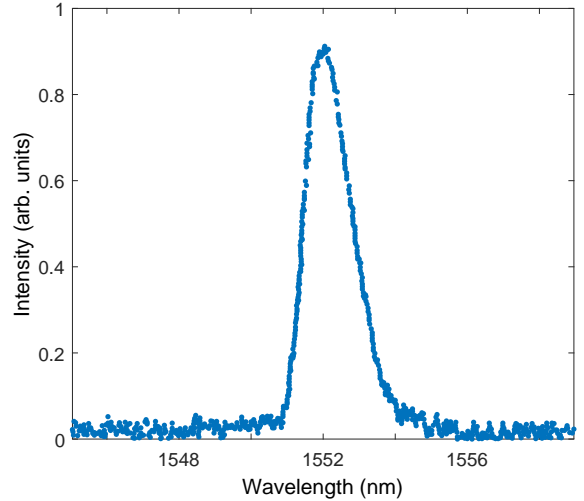}
\caption{Measured spectrum of the idler photons. The spectrum is
centered at 1552.3 nm with a bandwidth of 1.6 nm at full-width at half maximum (FWHM).}
\label{Spectrum1550}
\end{figure}

\section{Results}
Figure~\ref{PolCohLength} shows the measurement of the degree of
first-order coherence between signal photons, when the idler
photon generated in the first nonlinear crystal is reflected from
a mirror ($|\tau|=1$) that can be moved between two positions.
Note that this is a {\em proof-of-concept} experiment and the
axial resolution obtained (500 $\mu$m) can be readily improved,
as will be shown below in the discussion section.
The key element that makes our optical sectioning scheme work
is that the visibility of the interference of signal photons can
be controlled by insertion of different temporal delays between
the idler photons \cite{zou1993}. The curve measured, for any of
the two peaks, shows clearly a correlation function with a
triangular shape
\begin{eqnarray}
& & \left| g_{s_1,s_2}^{(1)}(T)\right|= \frac{\left|\langle
a_{s_1}^\dagger(t+T) a_{s_2}(t)\rangle \right|}{\sqrt{\langle
a_{s_1}^\dagger(t) a_{s_1}(t) \rangle} \sqrt{\langle
a_{s_2}^\dagger(t) a_{s_2}(t)\rangle}} \nonumber \\
& & =\text{tri} \left\{ \frac{1}{DL} \left[T-T_0\right] \right\},
\label{g1}
\end{eqnarray}
where 'tri' is the triangular function (see
the Appendix), $D$ is the difference of inverse group
velocities between signal and idler photons, $L$ is the nonlinear
crystal length, and $T_0$ ($cT_0$ spatial delay) is the
temporal delay between signal photons, necessary to obtain maximum
coherence between them (see the Appendix). The
temporal resolution is given by $DL$, which is the inverse
bandwidth of the source of photons and proportional to the
coherence length. The spatial resolution in free-space is thus
$cDL$, the axial resolution of our optical sectioning scheme.

\begin{figure}[t!] \centering
\includegraphics[width=8.6cm]{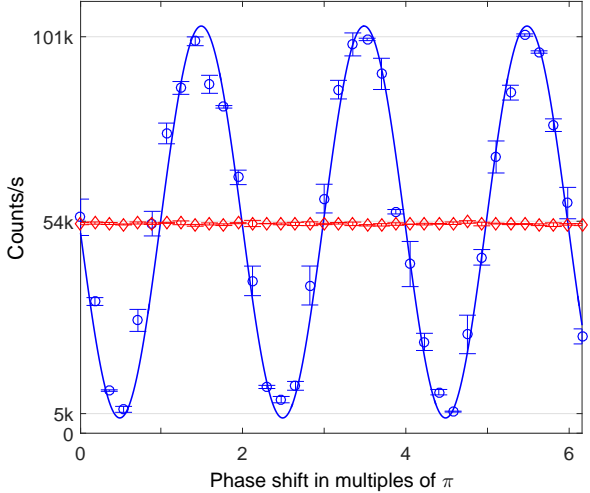}
\caption{Interference fringes for different values of the
reflection coefficient ($\tau$) measured with nanometric steps
within the red area marked in Fig.~\ref{PolCohLength}. Circles:
$|\tau|=1$; diamonds: $|\tau|=0$. The maximum visibility measured
is $V=90\%$. The error bars designate the standard deviation of
the experimental measures.} \label{PolFringes}
\end{figure}

The visibility of the left peak of the correlation function shown
in Fig.~\ref{PolCohLength}, is different than the one of the right
peak, with a value of $V=73\%$. The difference is caused by the
fact that signal coupling optimization was performed for one
location of the mirror, so when displaced, a small decrease of
visibility can be expected. In these results we made use of the
full bandwidth of the paired photons generated in both crystals,
changing the difference between path length of the two arms of the
interferometer in micrometric steps ($1 \, \mu$m).

\begin{figure}[t]
\centering
\includegraphics[width=8.6cm]{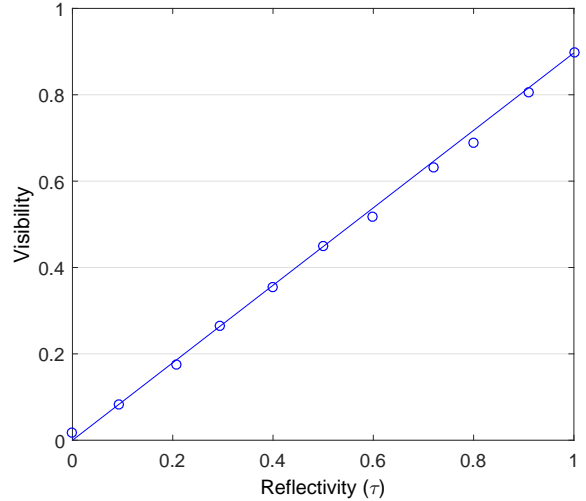}
\caption{Experimental relation between the visibility of the
interference pattern and the reflectivity of the simulated sample.
Open circles correspond to experimental measurements and the solid
curve stands for the theoretical prediction for our particular
visibility conditions, taking into account the characterized
transmission values of our neutral density filter.}
\label{PolVisRef}
\end{figure}

The coherence length and shape of the coherence functions are
directly related with the shape and bandwidth of the emitting
source, shown in Fig.~\ref{Spectrum1550}. In our case, the
20-mm-long PPLN type-0 crystal generates an idler spectral
emission bandwidth of about 1.6 nm at full-width at half maximum
(FWHM), measured with an OSA.
Measurements showed that the spectrum of photon pairs generated in
each crystal is slightly different. This is a source of spectral
distinguishability between photons coming from different crystals,
and therefore of loss of coherence and decrease of visibility.

In order to demonstrate that we are observing induced coherence in
the low parametric gain regime, we should obtain experimentally
the expected relationship between visibility and transmissivity,
i.e., $V=|\tau|$ (see Ref.~\cite{zou1991} and the Appendix).
Figures~\ref{PolFringes} and \ref{PolVisRef}, apart from showing
such relationship, also aim at demonstrating that one can obtain
the high visibility, $V=90\%$ in our case, necessary to reach high sensitivity when
mapping layer reflectivities. The erasure of distinguishability
between paired photons generated in different nonlinear crystals
is in general a highly demanding experimental task. In a recent
work, Barreto et al.~\cite{barreto2014} obtained a maximum
visibility of $V=77\%$. In the original paper from 1991 where the
idea of induced coherence was introduced by Zou et
al.~\cite{zou1991}, they were able to obtain a visibility of
$V=30\%$. These values from previous experiments show how
difficult it can be to successfully overlap spatial modes when large
bandwidths are considered, and to compensate all the different
degrees of freedom involved in the system that can provide
unwanted path distinguishability.

Figure~\ref{PolFringes} shows the number of signal photons detected at the output of PBS$_2$
with respect to variations in length of both arms of the
interferometer. Interference fringes appear for $|\tau|=1$. We
marked with a red rectangle in Fig.~\ref{PolCohLength}, the small
area that corresponds to the results shown in
Fig.~\ref{PolFringes}. We show (red diamonds) the effect of
blocking the first 1550 nm idler photon.

In these measurements the bandwidth of the signal photons is
filtered with the help of a 8-mm fiber Bragg grating (FBG). The
central wavelength of the FBG filters at room temperature is at
809.4 nm. This central wavelength can be modified by changing the
temperature of the FBG or stretching it, but we decided to change
the temperatures of the PPLN ovens instead. For that reason the
central wavelength of the SPDC idler spectrum in
Fig.~\ref{Spectrum1550} is around 1552.3 nm.

The purpose of filtering the signal photons with the FBG is
twofold. On the one hand, filtering out the bandwidth helps to
reduce the distinguishability of paired photons that originates in
different nonlinear crystals, erasing the undesired spectral
distinguishability. For that reason, the maximum visibility
measured in Fig.~\ref{PolFringes} increases up to $V=90\%$. On the
other hand, the coherence length turns out to be of the order of
tens of centimeters. This is due to the narrow bandwidth
($B_s~\sim~0.1$ nm) that is reflected from the FBG. Therefore
axial resolution degrades. Figure~\ref{PolVisRef} depicts the
experimental relationship between the visibility of the
interference pattern and the reflectivity of the sample. We note that all
results shown in this paper are raw experimental data with the
dark counts subtraction ($\sim 2$k) from the single-photon counting modules.

\section{Discussion and conclusions}
As a general rule in OCT, the broader the bandwidth of the
spectrum, the better the axial resolution. For the sake of
comparison, \emph{typical} OCT configurations available
commercially make use of broad-band light sources, with a spectral
emission of more than 100 nm at full-width at half maximum. In
this way they are able to perform measurements with axial
resolutions of the order of few microns, as shown in the inset of
Fig.~\ref{StandardOCT}. One can achieve
resolutions of a few nanometers (as narrow as 8
nm)~\cite{fuchs2017, fuchs2016} by using extreme ultraviolet
radiation. This would allow material identification and opaque
matter penetration, but would be harmful for biological
tissues.

We can  increase the resolution of the optical sectioning
scheme put forward here by increasing the bandwidth of
down-converted photons. Broader bandwidths can be obtained by
using shorter crystals, or by appropriately engineering the
phase-matching conditions of longer
crystals~\cite{hendrych2009,abolghasem2009}. In this way, axial
resolutions similar to the ones achieved with current OCT systems
are likely to be observed.

In conclusion, we have introduced the basic principles of an optical sectioning
imaging system that can be called {\em induced coherence
tomography}, a type of OCT scheme based on the concept of induced
coherence. We have demonstrated it using frequency-entangled
photons generated in SPDC processes embedded in a nonlinear
interferometer. Notwithstanding, Shapiro et
al.~\cite{shapiro2015} have shown that similar results can also be
obtained using a pair of bright pseudo-thermal beams possessing a
phase-sensitive cross correlation.

From a fundamental point of view, our scheme is a
coherence measurement, in contrast to {\em conventional} OCT
schemes that measure directly reflectivity. In our scheme, the
change of reflectivity induces a change of first-order coherence
between two streams of photons that are made to interfere. From a
practical point of view, the photons that are being measured never
interacts with the sample. That is to say, we are able to detect
photons centered at a wavelength with the maximum efficiency of
silicon based detectors, while the sample is being probed with
photons centered at NIR. This could potentially be used in
biological tissue to image even deeper into the tissue with
the use of longer wavelengths.

\section*{Acknowledgments}

We acknowledge financial support
from the Spanish Ministry of Economy and Competitiveness through
the Severo Ochoa Programme for Centres of Excellence in R$\&$D
(SEV- 2015-0522) and from Fundaci\'o Privada Cellex. A.V.
acknowledges support from the Claude Leon Foundation. J.P.T.
acknowledges support from the program ICREA Academia (Generalitat
de Catalunya). We thank N. Fleischman for her initial help in
building the optical set-up, and A. Barja for his collaboration
during the first stage of the design.

\widetext
\clearpage

\onecolumngrid

\section*{Appendix}

First-order coherence is the main tool in our {\em induced coherence
tomography} scheme. In this appendix we
calculate the value of the normalized first-order correlation
function between signal photons that are generated in different
nonlinear crystals, depicted in Fig.~\ref{PolCohLength}.

A CW plane-wave pump beam with frequency $\omega_p$ and flux of
pump photons of $F_0$ photons/s/m$^2$ illuminates a second-order
nonlinear crystal of length $L$ and nonlinear coefficient
$\chi^{(2)}$. The molecules or atoms of the crystal mediate the
generation of paired photons (signal and idler) by means of the
nonlinear process of spontaneous parametric down-conversion
(SPDC).

The electric field operators of signal and idler photons read
\begin{eqnarray}
& & E_{s}^{+} \left(t, z \right)= \frac{1}{\left(2 \pi
\right)^{1/2}} \int d\Omega \left[ \frac{\hbar \omega_s}{2
\epsilon_0 c n_s } \right]^{1/2} a_s \left(z,
\Omega \right)  \exp \left[ ik_s(\omega_s+\Omega) z-i \left( \omega_s+\Omega \right) t\right], \nonumber \\
& & E_{i}^{+} \left(t, z \right)= \frac{1}{\left(2 \pi
\right)^{1/2}} \int d\Omega \left[ \frac{\hbar \omega_i}{2
\epsilon_0 c n_i } \right]^{1/2} a_i \left(z, \Omega \right) \exp
\left[ ik_i (\omega_i+\Omega) z-i \left( \omega_i+\Omega \right)
t\right].
\end{eqnarray}
We neglect all spatial dependence of the fields for the sake of
simplicity.

\begin{figure}[t!]
\centering
\includegraphics[width=12cm]{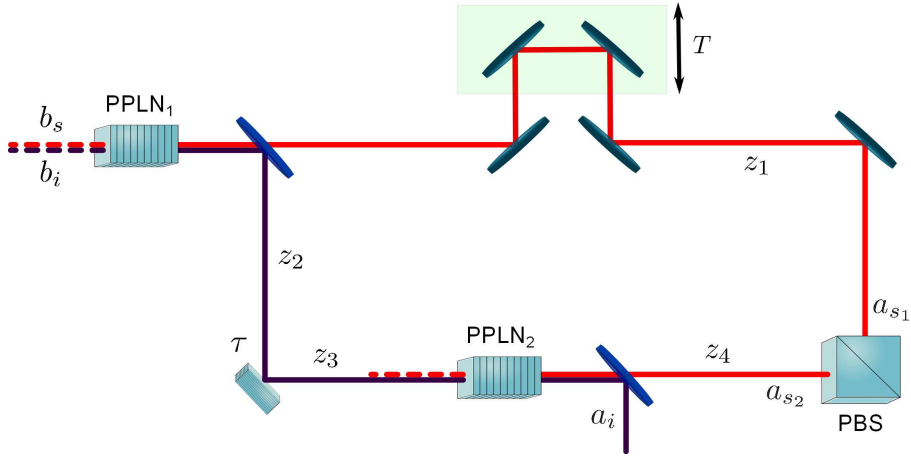}
\caption[Sketch of the experimental set-up]{Simple sketch of the
experimental set-up that shows the name of the operators at
different locations and main distances considered in the
calculation.} \label{schemeApp}
\end{figure}

Figure~\ref{schemeApp} shows schematically the operators at
different positions inside the experimental set-up, as well as the
main distances between elements considered. Let $\hat b_s(\Omega)
\equiv a_s(z=0,\Omega)$ and $\hat b_i(\Omega) \equiv a_i(z=0,
\Omega)$ designate the operators corresponding to the signal
(frequency $\omega_s+\Omega$) and idler (frequency
$\omega_i-\Omega$) modes at the input face of the nonlinear
crystal. $a_s(\Omega) \equiv a_s(z=L,\Omega)$ and $ a_i(\Omega)
\equiv a_i(z=L,\Omega)$ designate the operators corresponding to
the same modes at the output face of the nonlinear crystal. Under
the condition that the pump beam is undepleted, since the
efficiency of the parametric process is low, the relationship
between input and output modes is a Bogoliuvov transformation that
reads as~\cite{navez2001,brambilla2004,torres2011}
\begin{eqnarray}
& & a_s(\Omega) = U(\Omega) b_s(\Omega)  + V(\Omega)  b_i^\dagger(-\Omega) , \nonumber \\
& & a_i(\Omega) = U(\Omega) b_i(\Omega)   + V(\Omega)
b_s^\dagger(-\Omega),
\end{eqnarray}
where
\begin{eqnarray}
& & U(\Omega)=\exp \left[ i \frac{\Delta (\Omega) L}{2} \right] \left\{ \cosh \left[ \Gamma(\Omega) L \right] - i \frac{\Delta(\Omega)}{2\Gamma(\Omega)} \sinh \left[ \Gamma(\Omega)L \right] \right\}, \nonumber \\
& & V(\Omega)=i \frac{\sigma}{\Gamma(\Omega)} \exp \left[ i
\frac{\Delta (\Omega) L}{2} \right] \sinh \left[ \Gamma(\Omega)
\right] ,
\end{eqnarray}
$\sigma$ is the nonlinear coefficient (in units of $m^{-1}$)
\begin{equation}
\sigma= \left[ \frac{\hbar \omega_s \omega_i \omega_p \left[ \chi^{(2)} \right]^2 F_0 }{8 \epsilon_0 c^2 n_s n_i n_p} \right]^{1/2},
\end{equation}
the phase matching function is
\begin{equation}
\Delta(\Omega)= k_p(\omega_p)-k_s(\omega_s+\Omega_s)-k_i(\omega_i-\Omega),
\end{equation}
and $\Gamma$ reads
\begin{equation}
\Gamma(\Omega)= \left[\sigma^2-\frac{\Delta^2(\Omega)}{4} \right]^{1/2}.
\end{equation}

The nonlinear coefficient $\sigma$ is very small; therefore we can
safely write $\Gamma(\Omega) \sim i \Delta(\Omega)/2$. Moreover,
we expand the longitudinal wave numbers in a Taylor series, i.e.,
$k_j(\Omega)=k_j^0+N_j \Omega$ ($j=s,i$). $N_{s,i}$ are inverse
group velocities for the signal and idler photons.  The phase-matching function now reads $\Delta(\Omega)=D \Omega$ where
$D=N_i-N_s$. Under these conditions, we obtain that
\begin{eqnarray}
& & U(\Omega)=1, \nonumber \\
& & V(\Omega)=\sigma L\, \text{sinc} \left( \frac{DL \Omega}{2}
\right) \,\exp \left[ i \frac{\Delta (\Omega) L}{2} \right].
\end{eqnarray}

The normalized first-order correlation function
($g_{s_1,s_2}^{(1)}(T)$) between signal photons generated in the
first nonlinear crystal ($a_{s1}$) and signal photons generated in
the second nonlinear crystal ($a_{s2}$) (what we sometimes refer
to as {\em coherence} in the main text) writes
\begin{equation}
g_{s_1,s_2}^{(1)}(T) = \frac{\langle E_{s_1}^{-}(t+T)
E_{s_2}^{+}(t)\rangle}{\sqrt{\langle E_{s_1}^{-}(t) E_{s_1}^{+}(t)
\rangle} \sqrt{\langle E_{s_2}^{-}(t) E_{s_2}^{+}(t)\rangle}},
\end{equation}
where $T$ is a delay between signal photons traversing the upper
arm of the interferometer and signal photons traveling through the
lower arm.

At the sample, the operator transformation is \cite{haus2000,boyd2008}
\begin{equation}
a_i(\Omega) \Rightarrow \tau a_i(\Omega)+f(\Omega),
\end{equation}
where $\tau$ is the reflectivity of the sample and the commutation
relationship fulfills
$[f(\Omega),f^{\dagger}(\Omega)]=\delta(\omega-\omega^{\prime})$.
Taking this into account, the operators $a_{s_1}$, just before the
corresponding input port of the polarization beam splitter (see the
simple sketch in Fig.~\ref{schemeApp}), read
\begin{eqnarray}
& & \hspace{5mm}(i)\, b_s(\Omega) \nonumber \\
& & \Rightarrow (ii)\, \left[ U(\Omega) b_s(\Omega) + V(\Omega)
b_i^\dagger(-\Omega) \right] \exp \left[ i k_s (\Omega) L\right]
\nonumber \\
& & \Rightarrow (iii)\, a_{s_1}(\Omega)=\left[ U(\Omega)
b_s(\Omega) + V(\Omega) b_i^\dagger(-\Omega) \right]  \exp \left[
i k_s (\Omega) L+i k_s^{0}(\Omega) z_1\right]
\end{eqnarray}
The three expressions correspond to the operators (i) at the input
face of the nonlinear crystal, (ii) at the output face, and (iii)
before the polarizing beam splitter.

The operators $a_{s_2}$, just before the corresponding input port
of the polarization beam splitter, read
\begin{eqnarray}
& & \hspace{5mm}(i)\, b_{i}(\Omega) \nonumber \\
& & \Rightarrow  (ii)\, \left[ U(\Omega) b_i(\Omega) + V(\Omega)
b_s^{\dagger}(-\Omega) \right] \exp \left[ i k_i (-\Omega)
L\right] \nonumber \\
& & \Rightarrow  (iii)\,\left[ U(\Omega) b_i(\Omega) + V(\Omega)
b_s^{\dagger}(-\Omega) \right] \exp \left[ i k_i
(-\Omega) L+i k_i^0 (-\Omega) z_2\right] \nonumber \\
& & \Rightarrow  (iv)\, \tau \left[ U(\Omega) b_i(\Omega)  +
V(\Omega) b_s^{\dagger}(-\Omega) \right] \exp \left[ i k_i
(-\Omega)
L+i k_i^0i(-\Omega) z_2\right]+f(-\Omega) \nonumber \\
& & \Rightarrow  (v)\, \tau \left[ U(\Omega) b_i(\Omega)  +
V(\Omega) b_s^{\dagger}(-\Omega) \right] \exp \left[ i k_i
(-\Omega)
L+i k_i^0(-\Omega) (z_2+z_3)\right]+f(-\Omega) \exp \left[ i k_i^0(-\Omega) z_3 \right] \nonumber \\
& & \Rightarrow  (vi)\, U(\Omega) c_s(\Omega) \exp \left[ ik_s
(\Omega) L \right]+V(\Omega) \tau^* \left[ U^*(-\Omega)
b_i^{\dagger}(-\Omega) + V^*(-\Omega) b_s(\Omega) \right] \exp
\left[ ik_s(\Omega) L-i k_i (\Omega)
L-i k_i^0(\Omega) (z_2+z_3)\right] \nonumber \\
& & +V(\Omega) f^{\dagger}(\Omega) \exp \left[ ik_s(\Omega) L-ik_i^0 (\Omega) z_3 \right] \nonumber \\
& & \Rightarrow (vii)\, a_{s_2}= U(\Omega) c_s(\Omega) \exp \left[
ik_s(\Omega) L +i k_s^0
(\Omega) z_4 \right] \nonumber \\
& & + V(\Omega) \tau^* \left[ U^*(-\Omega) b_i^{\dagger}(-\Omega)
+ V^*(-\Omega) b_s(\Omega) \right] \exp \left[i k_s(\Omega) L -i
k_i (\Omega) L-i k_i^0(\Omega) (z_2+z_3)+i k_s^0(\Omega)
z_4\right] \nonumber
\\
& & +V(\Omega) f^{\dagger}(\Omega) \exp \left[ ik_s(\Omega)
L-ik_i^0 (\Omega) z_3+ik_s^0(\Omega) z_4 \right]
\end{eqnarray}
The six expressions correspond to the operators (i) at the input
face of the first nonlinear crystal, (ii) at the output face,
(iii) before the sample, (iv) after the sample, (v) before the
second nonlinear crystal, (vi) after the nonlinear crystal and
(vii) before the PBS.

The flux of photons of both signal beams are
\begin{equation}
N_{s_1}(t)= \langle a_{s_1}^{\dagger}(t) a_{s_1}(t) \rangle=
\frac{1}{2\pi}\int d\Omega |V(\Omega)|^2=\frac{(\sigma L)^2
}{2\pi}\int d\Omega \text{sinc}^2(\frac{DL
\Omega}{2})=\frac{\sigma^2 L}{D},
\end{equation}
and
\begin{equation}
N_{s_2}(t)= \langle a_{s_2}^{\dagger}(t) a_{s_2}(t) \rangle=
\frac{1}{2\pi}\int d\Omega |V(\Omega)|^2 \left[1+|\tau|^2
|V(\Omega)|^2 \right] \sim \frac{1}{2\pi}\int d\Omega
|V(\Omega)|^2 = \frac{\sigma^2 L}{D}.
\end{equation}

Under our experimental conditions, the flux of signal photons is
the same in both arms of the interferometer [see Fig. 2(c)].

The correlation function $|\langle a_{s_1}^{\dagger}(t) a_{s_2}(t)
\rangle|$ writes
\begin{eqnarray}
& & |\langle a_{s_1}^{\dagger}(t+T) a_{s_2}(t) \rangle|=
\frac{|\tau|}{2\pi} \left| \int d\Omega |V(\Omega)|^2 \exp \left[i
(\omega_s+\Omega)T -i k_i (\Omega) L-i k_i^0(\Omega) (z_2+z_3)+i
k_s^0(\Omega) z_4 \right]\right|
\nonumber \\
& & =\frac{|\tau|}{2\pi}  \int d\Omega \,\text{sinc}^2
\left(\frac{DL \Omega}{2} \right) \exp \left[i\frac{\Omega}{c}
\left(z_1+cT\right)-\left(z_4-c N_i L-z_2-z_3 \right) \right].
\end{eqnarray}

Therefore, we obtain
\begin{eqnarray}
& & \left| g_{s_1,s_2}^{(1)}(T)\right|= \frac{\left|\langle
a_{s_1}^\dagger(t+T) a_{s_2}(t)\rangle \right|}{\sqrt{\langle
a_{s_1}^\dagger(t) a_{s_1}(t) \rangle} \sqrt{\langle
a_{s_2}^\dagger(t) a_{s_2}(t)\rangle}} \nonumber \\
& & =\text{tri} \left\{ \frac{1}{cDL} \left[
\left(z_1+cT\right)-\left(z_4-c N_i L-z_2-z_3 \right) \right]
\right\},
\end{eqnarray}
where "tri" is the triangular function. This expression describes the result shown in Fig.~\ref{PolCohLength}, where one can see the triangular shape of the correlation function. The temporal resolution of the OCT scheme is given by $DL$, which is the inverse bandwidth of the source of photons. The spatial resolution in free space is thus $cDL$.

\vspace{0.35cm}
\twocolumngrid




\end{document}